\documentclass[aps,prd,twocolumn,groupedaddress,amssymb,eqsecnum,showpacs,epsfig]{revtex4}
\usepackage{graphicx}
\usepackage{bm}
\usepackage{dcolumn}
\usepackage{amsmath}
\def \lleq {\lower0.9ex\hbox{ $\buildrel < \over \sim$} ~}
\def \ggeq {\lower0.9ex\hbox{ $\buildrel > \over \sim$} ~}

\def \omm  {\Omega_{0 {\rm m}}}
\def \oml   {\Omega_{\Lambda}}

\def \beq  {\begin{equation}}
\def \eeq  {\end{equation}}
\def \ber  {\begin{eqnarray}}
\def \eer  {\end{eqnarray}}

\def\apj{{Astroph.\@ J.\ }}

\def\aj{{Astron.\@ J.\ }}

\begin{document}
\newcommand{\newc}{\newcommand}

\newc{\be}{\begin{equation}}
\newc{\ee}{\end{equation}}
\newc{\ba}{\begin{eqnarray}}
\newc{\ea}{\end{eqnarray}}
\newc{\bea}{\begin{eqnarray*}}
\newc{\eea}{\end{eqnarray*}}
\newc{\D}{\partial}
\newc{\ie}{{\it i.e.} }
\newc{\eg}{{\it e.g.} }
\newc{\etc}{{\it etc.} }
\newc{\etal}{{\it et al.}}
\newcommand{\nn}{\nonumber}
\newc{\ra}{\rightarrow}
\newc{\lra}{\leftrightarrow}
\newc{\lsim}{\buildrel{<}\over{\sim}}
\newc{\gsim}{\buildrel{>}\over{\sim}}
\title{The Limits of Extended Quintessence}
\author{S. Nesseris$^a$ and L. Perivolaropoulos$^b$ }
\affiliation{Department of Physics, University of Ioannina, Greece
\\ $^a$ e-mail: me01629@cc.uoi.gr, $^b$ e-mail:
leandros@uoi.gr}
\date {\today}

\begin{abstract}
We use a low redshift expansion of the cosmological equations of
extended (scalar-tensor) quintessence to divide the observable
Hubble history parameter space in four sectors:  A forbidden sector
I where the scalar field of the theory becomes imaginary (the
kinetic term becomes negative), a forbidden sector II where the
scalar field rolls up (instead of down) its potential, an allowed
`freezing' quintessence sector III where the scalar field is
currently decelerating down its potential towards freezing and an
allowed  `thawing' sector IV where the scalar field is currently
accelerating down its potential. The dividing lines between the
sectors depend sensitively on the time derivatives of the Newton's
constant $G$ over powers of the Hubble parameter. For minimally
coupled quintessence which appears as a special case for a constant
$G$ our results are consistent with previous studies. Observable
parameter $\chi^2$ contours based on current data (SNLS dataset) are
also constructed on top of the sectors, for a prior of
$\Omega_m=0.24$. By demanding that the observed $2\sigma$ $\chi^2$
parameter contours do not lie entirely in the forbidden sectors we
derive stringent constraints on the current second time derivative
of Newton's constant $G$. In particular we find $\frac{\ddot G}{G
}>-1.91\; H_0^2=-2\times 10^{-20}h^2\;yrs^{-2}$ at the $2\sigma$
level which is complementary to solar system tests which constrain
only the first derivative of $G$ as $|\frac{\dot
G}{G}|<10^{-14}yrs^{-1}$ at $1\sigma$.
\end{abstract}
\maketitle

\section{Introduction}

Cosmological observations based mainly on type Ia supernovae (SnIa)
standard candles \cite{Astier:2005qq,Riess:2004nr,snobs} and also on
standard rulers \cite{Allen:2004cd,Eisenstein:2005su,Spergel:2006hy}
have provided a fairly accurate form of the Hubble parameter
$H(z)=\frac{\dot a}{a}$ as a function of redshift $z$ in the
redshift range $0<z<1.7$ \cite{Riess:2004nr} and beyond
\cite{Spergel:2006hy}. This form indicates that despite the
attractive gravitational properties of matter, the universe has
entered a phase of accelerated expansion at a redshift $z\simeq
0.5$. It is therefore clear that the simplest cosmological model,
where the universe is dominated by matter and its dynamics is
determined by general relativity is ruled out at several $\sigma$
\cite{snobs}. The central current questions in cosmology research
are the following:
\begin{itemize} \item What theoretical models are
consistent with the currently detected form of $H(z)$? \item What
are the generic predictions of these models with respect to $H(z)$
so that they can be ruled out or confirmed by more detailed
observations of $H(z)$? \end{itemize}

In a class of approaches the required gravitational properties of
dark energy (see
\cite{Copeland:2006wr,Padmanabhan:2006ag,Perivolaropoulos:2006ce,Sahni:2006pa}
for recent reviews) needed to induce the accelerating expansion
are well described by its equation of state
$w(z)=\frac{p_X(z)}{\rho_X(z)}$. In this case the simplest model
consistent with the currently detected form of $H(z)$ is the flat
$\Lambda CDM$ model. According to this model, the universe is flat
and its evolution is determined by general relativity with a
cosmological constant through the Friedman equation \be H(z)^2 =
H_0^2 (\omm (1+z)^3 + \oml) \ee where $\omm=\frac{\rho_m}{\rho_c}$
is the current matter density normalized on the critical density
for flatness $\rho_c$ and $\oml=1-\omm$ is a constant density due
to the cosmological constant. The main advantages of this model
are {\it simplicity} and {\it predictability}: It has a single
free parameter and it can be definitively ruled out by future
observations. Its disadvantages are {\it lack of theoretical
motivation} and {\it fine tuning}: There is no physically
motivated theoretical model predicting generically a cosmological
constant at the observed value. This value is $120$ orders of
magnitude smaller than its theoretically expected value
\cite{Carroll:2000fy,Sahni:1999gb}.

Attempts to replace the cosmological constant by a minimally coupled
dynamical scalar field (minimally coupled quintessence (MCQ)
\cite{Ratra:1987rm,Caldwell:1997ii,Zlatev:1998tr}) have led to
models with a vastly larger number of parameters fueled by the
arbitrariness of the scalar field potential. Despite this
arbitrariness and vast parameter space, quintessence models are
generically constrained to predicting a limited range of functional
forms for $H(z)$ \cite{Boisseau:2000pr,Perivolaropoulos:2005yv}.
This is a welcome feature which provides ways to either rule out or
confirm this class of theories.

The allowed functional space of $H(z)$ can be further increased by
considering models based on extensions of general relativity such as
braneworlds
\cite{Sahni:2002dx,Bogdanos:2006pf,Kofinas:2005hc,Dvali:2000rv},
$f(R)$ theories \cite{Nojiri:2006gh} or scalar-tensor theories
\cite{Esposito-Farese:2000ij,Boisseau:2000pr,Perivolaropoulos:2005yv}
(extended quintessence (EXQ) \cite{Perrotta:1999am}) where the
accelerated expansion of the universe is provided by a non-minimally
coupled scalar field. This class of theories is strongly motivated
theoretically as it is predicted by all theories that attempt to
quantize gravity and unify it with the other interactions. On the
other hand, its parameter space is even larger than the
corresponding space of MCQ since the later is a special case of EXQ.
Local (eg solar system) gravitational experiments and cosmological
observations constrain the allowed parameter space to be close to
general relativity. Despite of these constraints however, the
allowed by EXQ functional forms of $H(z)$ are significantly more
than those allowed by MCQ. The detailed identification of the
forbidden $H(z)$ functional forms for both MCQ and EXQ is
particularly important since it may allow future observations
determining $H(z)$ to rule out one or both of these theories.

Previous studies \cite{Caldwell:2005tm} have mainly focused on the
$H(z)$ limits of MCQ using a combination of plausibility arguments
and numerical simulations of several classes of potentials. The low
redshift $H(z)$ parameter space was divided in three sectors: a
forbidden sector which could not correspond to any plausible
quintessence model, a sector corresponding to the {\it freezing}
quintessence scenario and a sector corresponding to the {\it
thawing} quintessence scenario. In the freezing quintessence models,
a field $\Phi$ which was already rolling towards its potential
minimum prior to the onset of acceleration slows down (${\ddot
\Phi}<0$) and creeps to a halt (freezes) mimicking a cosmological
constant as it comes to dominate the universe. In the thawing
quintessence models, the field has been initially halted by Hubble
damping at a value displaced from its minimum until recently when it
`thaws' and starts to roll down to the minimum (${\ddot \Phi}>0$).

Here we extend these studies to the case of EXQ. Instead of using
numerical simulations however, applied to specific potential
classes, we use generic arguments demanding only the internal
consistency of the theory. Thus our `forbidden' sector when reduced
to MCQ is smaller but more generic (applicable to a more general
class of models) than that of Ref. \cite{Caldwell:2005tm} (see also
Refs. \cite{Scherrer:2005je,Chiba:2005tj}).

The size and location of the sectors of the low $z$, $H(z)$
parameter space, depends sensitively on the assumed current time
derivatives of the Newton's constant $G(t)$ and reduce to well known
results in the MCQ limit of $G(t)=G_0=const$. Therefore an
interesting interplay develops between local gravitational
experimental constraints of the current time derivatives of $G(t)$
(eg $\frac{\dot G_0}{G_0}$ or $\frac{\ddot G_0}{G_0}$) and
cosmological observations of $H(z)$ at low redshifts. For example, a
constraint on $\frac{\dot G_0}{G_0}-\frac{\ddot G_0}{G_0}$ from
local gravitational experiments defines the forbidden sector in the
low $z$ expansion coefficients of $H(z)^2$ in the context of EXQ. If
such coefficients are measured to be in the forbidden sector by
cosmological observations then EXQ could be ruled out.
Alternatively, if such coefficients are measured to be in the
forbidden sector for MCQ but in the allowed sector of EXQ (either
`freezing' or `thawing') then this would rule out MCQ in favor of
EXQ. As shown in what follows, current observational constraints on
$H(z)$ imply significant overlap with the allowed sectors of both
MCQ and EXQ. This however may well change in the near future with
more accurate determinations of $H(z)$ and the time derivatives of
$G_0$.

\section{The Boundaries of Extended Quintessence}

Extended quintessence is based on the simplest but very general
(given its simplicity) extension of general relativity: {\it
Scalar-Tensor theories}. In these theories Newton's constant obtains
dynamical properties expressed through the potential $F(\Phi)$. The
dynamics are determined by the Lagrangian density
\cite{Boisseau:2000pr,Esposito-Farese:2000ij} \be {\cal
L}=\frac{F(\Phi)}{2}~R - \frac{1}{2}~g^{\mu\nu}
\partial_{\mu}\Phi
\partial_{\nu}\Phi
- U(\Phi)  + {\cal L}_m[\psi_m; g_{\mu\nu}]\  \label{lst} \ee where
${\cal L}_m[\psi_m; g_{\mu\nu}]$ represents matter fields
approximated by a pressureless perfect fluid. The function $F(\Phi)$
is observationally constrained as follows:
\begin{itemize}
\item $F(\Phi)>0$ so that gravitons carry positive
energy\cite{Esposito-Farese:2000ij}. \item
$\frac{1}{F}\left(\frac{dF}{d\Phi}\right)^2\vert_{z=0}<\frac{1}{4}10^{-4}$
from solar system observations \cite{will-bounds}.
\end{itemize}
In such a model the effective Newton's constant for the attraction
between two test masses is given by \be G_{eff}(t)
 =\frac{1}{F(t)}\frac{F(t)+2\left(\frac{dF}{d\Phi}(t)\right)^2}{F(t)+\frac{3}{2}\left(\frac{dF}{d\Phi}(t)\right)^2}
 \simeq \frac{1}{F(t)}=G(t) \label{gf} \ee
where the approximation of equation (\ref{gf}) is valid at low
redshifts. Assuming a homogeneous $\Phi$ and varying the action
corresponding to (\ref{lst}) in a background of a flat FRW metric,
we find the coupled system of generalized Friedman equations \ba
3F H^2 &=&
\rho + \frac{1}{2} {\dot \Phi}^2 -3H{\dot F} + U \label{stfe1} \\
-2F{\dot H} & = & \rho + p + {\dot \Phi}^2 + {\ddot F} - H{\dot F}
\label{stfe2} \ea where we have assumed the presence of a perfect
fluid $(\rho=\rho_m, p\simeq 0)$ playing the role of matter fields.
Expressing in terms of redshift and eliminating the potential $U$
from equations (\ref{stfe1}), (\ref{stfe2}) we find
\cite{Esposito-Farese:2000ij,Perivolaropoulos:2005yv} \ba
&&\Phi'^2=-F''-\left[(lnH)'+ \frac{2}{1+z}\right]F' + 2
\frac{(lnH)'}{1+z}F - \nn
\\&&-3(1+z)\omm \left(\frac{H_0}{H}\right)^2 F_0
\label{stphp} \ea

\noindent where $'$ denotes derivative with respect to redshift and
$F_0$ is set to 1 in units of $\frac{1}{8\pi G_0}$ and corresponds
to the present value of $F$.  Alternatively, expressing in terms of
redshift and eliminating the kinetic term $\Phi'^2$ from equations
(\ref{stfe1}), (\ref{stfe2}) we find \ba U&=&\frac{(1+z)^2 H^2}{2}[
F'' + \left[(\ln H)'-\frac{4}{1+z}\right]F' + \nn \\ &+&
\left[\frac{6}{(1+z)^2} - \frac{2}{1+z}(\ln H)'\right]~F - \nn \\
&-& 3 (1+z) \left({H_0\over H}\right)^2 F_0  \Omega_{m,0}]
\label{stu} \ea

 We now wish to explore the observational
consequences that emerge from the following generic inequalities
anticipated on
a purely theoretical level \ba \Phi'(z)^2 &>& 0 \label{in1}\\
U'(z)&>&0 \label{in2}\\
(\Phi'(z)^2)'&>&0 \;\;\; (freezing) \label{in3} \\
(\Phi'(z)^2)'&<&0 \;\;\; (thawing)\label{in4} \ea

The inequality (\ref{in1}) is generic and merely states that the
scalar field in scalar-tensor theories is real as it is directly
connected to an observable quantity (Newton's constant). The
inequality (\ref{in2}) is also very general as it merely states
that the scalar field rolls down (not up) its potential. This
inequality is not as generic as (\ref{in1}) since it implicitly
assumes a monotonic potential. Finally, the inequality (\ref{in3})
((\ref{in4})) denotes a scalar field which decelerates
(accelerates) as it rolls down its potential thus corresponding to
a freezing (thawing) quintessence model.

Since we are interested in the observational consequences of
equations (\ref{in1})-(\ref{in4}) at low redshifts, we consider
expansions of equations (\ref{stphp}) and (\ref{stu}) around $z=0$
expanding $F(z)$, $H(z)^2$ $U(z)$, $\Phi(z)$ as follows 
\cite{Gannouji:2006jm}: \ba F(z)&=& 1+F_1 z + F_2
z^2 + ... \label{fexp} \\ H(z)^2 &=& 1 +h_1 z + h_2 z^2 + ... \label{hexp} \\
U(z)&=&1+U_1 z + U_2 z^2 + ... \label{uexp} \\
\Phi(z) &=& 1+ \Phi_1 z + \Phi_2 z^2 + ... \label{pexp} \ea where
we have implicitly normalized over $F_0$, $H_0$, $U_0$ and
$\Phi_0$.

It is straightforward to connect the expansion coefficients $F_i$
of equation (\ref{fexp}) with the current time derivatives of
$G(t)$ using equation (\ref{gf}) and the time-redshift relation
\be \frac{dt}{dz}=-\frac{1}{H(z)(1+z)} \label{tz} \ee For example
for $F_1$ we have \be
F_1=\frac{1}{F_0}\frac{dF}{dz}\vert_{z=0}=\frac{{\dot
G}_0}{G_0H_0}\equiv g_1 \label{g1d} \ee where the subscript $_0$
denotes the present time and $H_0\simeq 10^{-10}\;h\;yrs^{-1}$.
Similarly, for $F_2$ we find \be
F_2=g_1(g_1-\frac{h_1+2}{4})-\frac{g_2}{2} \label{g2d} \ee where
we have defined \be g_n\equiv \frac{G_0^{(n)}}{G_0 H_0^n}
\label{gnd} \ee with the superscript $^{(n)}$ denoting the
$n^{th}$ time derivative. We may now substitute the expansions
(\ref{fexp}), (\ref{hexp}), (\ref{pexp}) in equation (\ref{stphp})
replacing the coefficients $F_i$ by the appropriate combination of
$g_n$. Equating terms order by order in $z$ and ignoring terms
proportional to $g_1$ due to solar system constraints which imply
\cite{pitjeva,Muller:2005sr} \be \vert g_1 \vert <10^{-13}
yrs^{-1}H_0^{-1}\simeq 10^{-3}h^{-1} \ll 1 \label{g1c} \ee we find
for the zeroth and first order
in $z$ \ba &h_1-3\omm +g_2 = \Phi_1^2 > 0&  \label{hc1} \\
&-h_1(1+h_1)+2h_2-3\omm (1-h_1)-  & \nn \\&-g_2 (1+h_1)-g_3=4\Phi_1
\Phi_2=(\Phi^{\prime 2})^\prime(z=0)& \label{hc2} \ea The inequality
(\ref{hc1}) defines a forbidden sector for extended quintessence for
each value of $g_2$. For $g_2=0$ this reduces to the well known
result that MCQ can not cross the phantom divide line (see equation
(\ref{wc1}) below). Equation (\ref{hc2}) can be used along with
(\ref{in3}), (\ref{in4}) to divide the allowed $(h_1,h_2)$ parameter
sector into a freezing quintessence sector $(\Phi'(z)^2)'>0$ and a
thawing quintessence sector $(\Phi'(z)^2)'<0$ for each set of
$(g_2,g_3)$.

Unfortunately, solar system gravitational experiments have so far
provided constraints for $g_1$ \cite{pitjeva} (equation (\ref{g1c}))
but not for $g_i$ with $i\geq 2$. This lack of constraints is not
due to lack of observational data quality but simply due to the fact
that existing codes have parameterized $G(t)$ in the simplest
possible way ie as a linear function of $t$. It is therefore
straightforward to extend this parameterizations to include more
parameters thus obtaining constraints of $g_i$ with $i\geq 2$.
\begin{figure*}
\rotatebox{0}{\resizebox{1\textwidth}{!}{\includegraphics{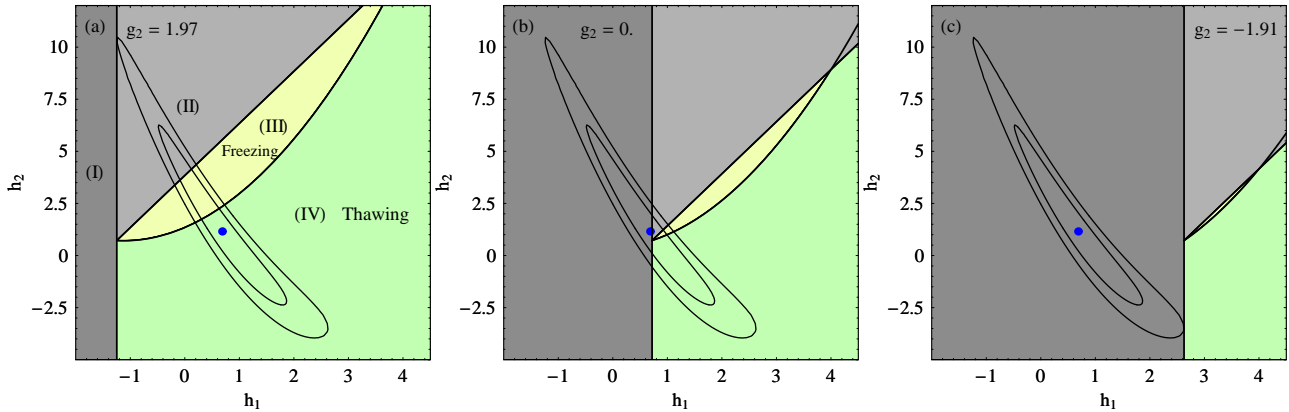}}}
\vspace{0pt}{ \caption{The $h_1-h_2$ sectors of EXQ: Sector I is the
forbidden sector where the scalar field becomes imaginary. Sector II
is also forbidden and corresponds to a scalar field that rolls up
its potential. Sector III corresponds to `freezing' EXQ where the
field is decelerating down its potential towards freezing. Sector IV
corresponds to `thawing' EXQ where the scalar field is accelerating
down its potential. The forbidden sector I shrinks for a Newton's
constant $G$ that decreases with time ($g_2 >0$) while the `freezing
sector' expands (Fig. 1a). The boundaries of the sectors are
provided by (\ref{hc1}),(\ref{hc2}) and (\ref{hc3}). The $1\sigma$,
$2\sigma$ $\chi^2$ contours obtained from the SNLS SnIa dataset for
$\omm=0.24$ using the CPL parameterization, are also shown.}}
\label{fig1}
\end{figure*}
Such an analysis is currently in progress \cite{pitjeva1} but even
before the results become available we can use some heuristic
arguments to estimate the order of magnitude of the expected
constraints on $g_i$ ($i\geq 2$) given the current solar system
data.

Current solar system gravity experiments are utilizing lunar laser
ranging \cite{Muller:2005sr} and high precision planet ephemerides
data \cite{pitjeva} to compare the trajectories of celestial
bodies with those predicted by general relativity. The possible
deviation from the general relativity predictions is parameterized
\cite{Muller:2005sr} using the PPN parameters $\beta$ and
$\gamma$, the first current derivative of Newton's constant
$\frac{{\dot G}_0}{G_0}$ and a Yukawa coupling correction to
Newton's inverse square law. These experiments have been
collecting data for a time $\Delta t$ of several decades
\cite{Muller:2005sr,pitjeva} ie $\Delta t = {\cal O}(100yrs)$. The
current $1\sigma$ constraint \cite{pitjeva} of \be \vert
\frac{{\dot G}_0}{G_0}\vert=\vert -0.2 \pm 0.5 \vert \times
10^{-13}yrs^{-1}<10^{-13}yrs^{-1} \ee implies an upper bound on
the total variation $\frac{\Delta G}{G_0}$ over the time-scale
$\Delta t$ of approximately \be \vert \frac{\Delta G}{G_0} \vert
\simeq \vert \frac{{\dot G}_0}{G_0} \vert \Delta t < 10^{-11}
\label{dglim1}\ee The same bound is obtained by considering the
relative error in the orbital periods $T$ of the Earth and other
planets which at $1\sigma$ is \cite{pitjeva} \be \frac{\Delta
T}{T}<10^{-12} \label{dtc} \ee Given that the Keplerian orbital
period is \be T_K=\sqrt{\frac{4\pi^2 r^3}{Gm}}\sim
G^{-\frac{1}{2}} \label{kep} \ee we find \be \vert \frac{\Delta
T}{T}\vert =\frac{1}{2}\vert \frac{\Delta G}{G_0}\vert < 10^{-12}
\label{dgc1} \ee in rough agreement with (\ref{dglim1}).

Using the upper bound (\ref{dglim1}) and attributing any variation
of $G$ to a term quadratic in $\Delta t$ we get \be \vert
\frac{\Delta G}{G_0} \vert \simeq \vert \frac{{\ddot G}_0}{G_0}
\vert (\Delta t)^2 < 10^{-11} \label{dglim2}\ee which implies that
\be \vert \frac{{\ddot G}_0}{G_0} \vert < 10^{-15} yrs^{-2} 
\implies \vert g_2 \vert < 10^5 h^{-2} \label{dglim3}\ee giving a
rough order of magnitude estimate of the upcoming constraints on
$g_2$. 
Preliminary results from the analysis of solar system data
indicate that \cite{privateMuller} \be \frac{{\ddot G}_0}{G_0}
\simeq (4\pm 5) \cdot 10^{-15} yrs^{-2} \label{mulbound}\ee which
is not far off the rough estimate of equation (\ref{dglim3}).
Generalizing the above arguments to arbitrary order in $\Delta t$
we find \be \vert g_n \vert < 10^{8n-11}h^{-n} \label{gnlim}\ee
which implies that current solar system tests can not constrain
$g_n$ in any cosmologically useful way for $n\geq 2$.

We now return to the constraint equations (\ref{hc1}) and
(\ref{hc2}) and supplement them by the constraint obtained from the
inequality (\ref{in2}). Using the expansions (\ref{fexp}),
(\ref{hexp}) and (\ref{uexp}) in equation (\ref{stu}) and equating
terms of first order in $z$ we find \be
U_1=U'(z=0)=\frac{1}{2}(5h_1-2h_2-9 \omm +5g_2+g_3)>0 \label{hc3}
\ee where as usual we have ignored terms proportional to $g_1$ due
to (\ref{g1c}). We now may use (\ref{hc1}), (\ref{hc2}) and
(\ref{hc3}) to define the following sectors in the $h_1-h_2$
parameter space for fixed $g_1$, $g_2$: \begin{itemize} \item A
forbidden sector I where the inequality (\ref{hc1}) is violated.
\item A forbidden sector II where the inequality (\ref{hc3}) is
violated but not the inequality (\ref{hc1}). \item An allowed sector
III of freezing EXQ where the inequalities (\ref{in1}), (\ref{in2})
and (\ref{in3}) are respected while (\ref{in4}) is violated. \item
An allowed sector IV of thawing EXQ where the inequalities
(\ref{in1}), (\ref{in2}) and (\ref{in4}) are respected while
(\ref{in3}) is violated. \end{itemize}
\begin{figure*}
\rotatebox{0}{\resizebox{1\textwidth}{!}{\includegraphics{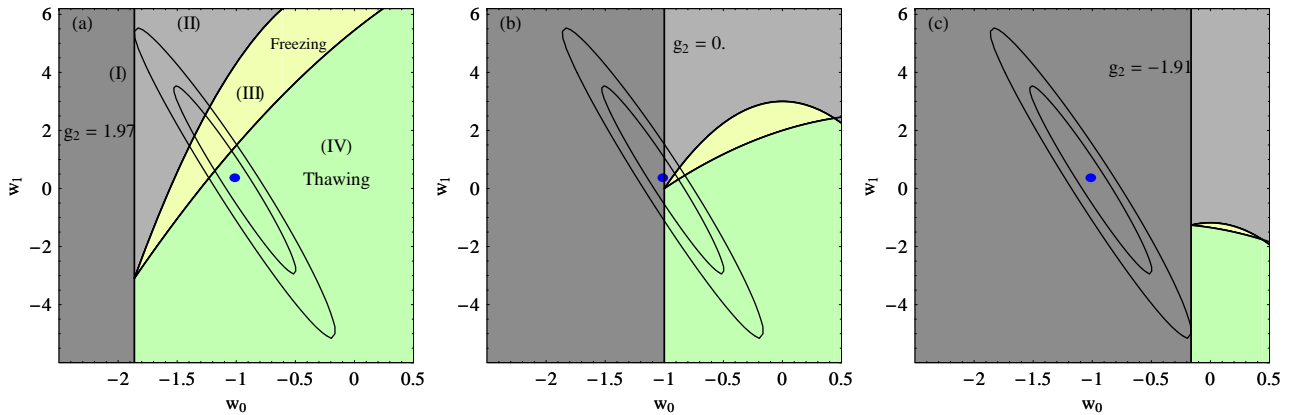}}}
\vspace{0pt}{ \caption{The  sectors I-IV of Fig. 1 mapped on the
$w_0-w_1$ parameter space. The allowed sectors extend to $w_0<-1$
for $g_2>0$.}} \label{fig2}
\end{figure*}
These sectors are shown in Fig. 1 for
$\omm=0.24$\cite{Spergel:2006hy}, $g_3=0$ and three values of $g_2$:
$g_2=1.97$, $g_2=0$ and $g_2=-1.91$.

It would be interesting to superpose on these sectors of Fig. 1 the
parameter $\chi^2$ contours obtained by fitting to the SnIa data. It
is not legitimate however to extrapolate the expansions in powers of
$z$ out to $z={\cal O}(1)$ where the SnIa data extend. We thus use
an extrapolation\cite{Chevallier:2000qy,Linder:2002et} with several
useful properties first proposed by Chevallier and
Polarski\cite{Chevallier:2000qy} and later investigated in more
detail by Linder\cite{Linder:2002et}. We superpose on the sectors of
Fig. 1, the $1\sigma$ and $2\sigma$ $h_1-h_2$ $\chi^2$ contours
obtained by fitting the Chevalier-Polarski-Linder (CPL)
\cite{Chevallier:2000qy,Linder:2002et} parametrization \ba H^2
(z)&=&H_0^2 [ \Omega_{0m} (1+z)^3 + \nn
\\ &+& (1-\Omega_{0m})(1+z)^{3(1+w_0+w_1)}e^{\frac{-3w_1 z}{(1+z)}}]
\label{hcpl} \ea to the SNLS dataset \cite{Astier:2005qq} with the
same $\omm$ prior ($\omm=0.24$). Using equation (\ref{hcpl}) along
with the expansion (\ref{hexp}) it is easy to
show that \ba h_1 &=& 3(1+w_0 - \omm w_0) \label{hiwi1}\\
h_2 &=& \frac{3}{2} (2 + 5 w_0 (1-\omm) + \nn \\ &+& 
(1-\omm)(3w_0^2 + w_1)) \label{hiwi2} \ea so that the standard
$\chi^2$ contours in the $w_0 - w_1$ space (see eg
\cite{Nesseris:2005ur}) can be easily translated to the $h_1 -
h_2$ space of Fig. 1. The CPL parametrization (\ref{hcpl}) is
constructed so that the parameter $w(z)$ defined as \be
w(z)=\frac{\frac{2}{3}(1+z)\frac{d ln H}{dz}
-1}{1-\frac{H_0^2}{H^2} \omm (1+z)^3} \label{wzh1} \ee to be of
the form \be w(z)=w_0 + w_1 \frac{z}{1+z} \label{cplpar} \ee which
has useful and physically motivated
properties\cite{Linder:2002et}. The parameter $w(z)$ is
particularly useful and physically relevant in the MCQ limit of
$g_i \rightarrow 0$. In that limit $w(z)$ becomes the MCQ equation
of state parameter ie \be
w(z)=\frac{p_{MCQ}}{\rho_{MCQ}}=\frac{\frac{1}{2}{\dot \Phi}^2 -
U(\Phi)}{\frac{1}{2}{\dot \Phi}^2 + U(\Phi)} \label{wzmc} \ee as
may be shown from the MCQ Friedman equations. The values of $g_2$
used in Figs 1a and 1c were motivated by demanding minimum and
maximum overlap of the forbidden sector I with the $2\sigma$ $h_1
- h_2$ contour.

The following comments can be made with respect to Fig. 1:
\begin{itemize} \item For values of $g_2 < -1.91$, the $2\sigma$
parameter contour obtained from SNLS lie entirely in the forbidden
sector. This bound is independent of $g_3$ which does not enter in
the inequality (\ref{hc1}). Thus in the context of EXQ the
constraint on $g_2$ obtained by the SNLS data at $2\sigma$ level
is \be g_2=\frac{{\ddot G}_0}{G_0 H_0^2}>-1.91 \label{ddgc} \ee
Notice the dramatic improvement of this constraint (with respect
to the lower bound) compared to the anticipated constraint of
$-10^5< g_2 < 10^5$ anticipated from solar system data (equation
(\ref{dglim3}))! \item The parameter sector III of freezing
quintessence is significantly smaller than sector IV of thawing
quintessence and the difference is more prominent for smaller
$g_2$. \item For $g_2>0$ the allowed parameter space increases
significantly compared to MCQ ($g_2=0$). Therefore, if future
cosmological observations show preference to the forbidden sectors
I or II of Fig. 1b (MCQ) this could be interpreted as evidence for
EXQ with $g_2>0$ (see also comments in Refs.
\cite{Perivolaropoulos:2005yv,Nesseris:2006er,Gannouji:2006jm})
\end{itemize}

Even though the plots of Fig. 1 capture the full physical content of
our results it is useful to express the sectors I-IV in terms of
parameter pairs other than $h_1-h_2$ which are more common in the
literature. Such parameters are the expansion coefficients $w_i$ of
$w(z)$ \be w(z)=w_0+w_1 z + w_2 z^2 + ... \label{wexp} \ee which is
connected to $H(z)$ via equation (\ref{wzh1}). By expanding both
sides of equation (\ref{wzh1}) using equations (\ref{hexp}) and
(\ref{wexp}) we may express $h_i$ in terms of $w_i$ thus rederiving
equations (\ref{hiwi1}) and (\ref{hiwi2}) for $i=0$ and $i=1$. The
result is identical since the $w_0$, $w_1$ expansion coefficients of
equation (\ref{wexp}) coincide with the $w_0$, $w_1$ coefficients of
the CPL parametrization for a $w(z)$ given by equation
(\ref{cplpar}). The advantage of using the parameters $w_0-w_1$
instead of $h_1-h_2$ is that they provide better contact with
previous studies and can illustrate clearly the fact that a $g_2>0$
can provide a phantom behavior $w_0<-1$ and crossing of the phantom
divide line $w=-1$ in EXQ models.
\begin{figure*}
\rotatebox{0}{\resizebox{1\textwidth}{!}{\includegraphics{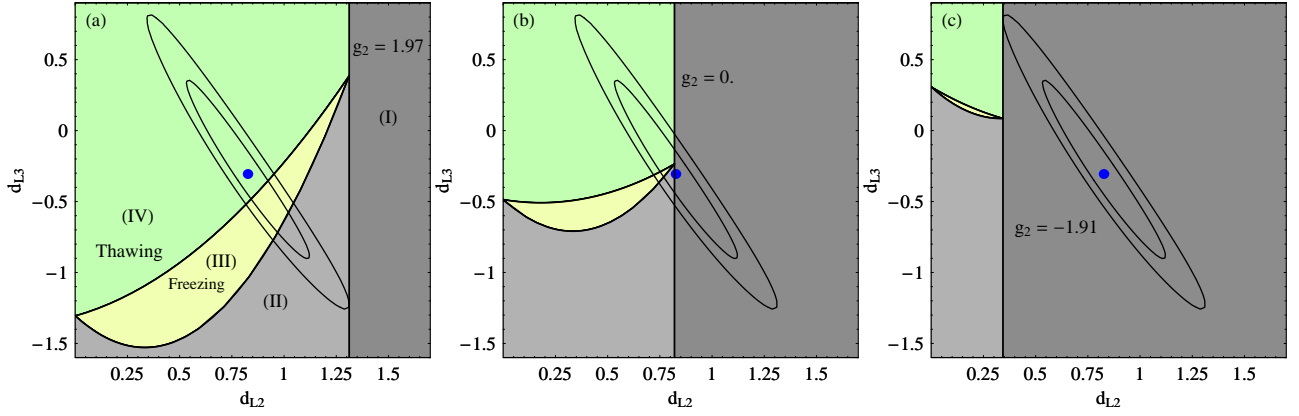}}}
\vspace{0pt}{ \caption{The  sectors I-IV of Fig. 1 mapped on the
$d_{L2}-d_{L3}$ parameter space. Notice that in this case the
forbidden sector I is on the right (large $d_{L2}$).}} \label{fig3}
\end{figure*}
Using equations (\ref{hiwi1}) and (\ref{hiwi2}) we may express the
constraint equations (\ref{hc1}), (\ref{hc2}) and (\ref{hc3}) that
define the sectors I-IV in Fig. 1 in terms of $w_0-w_1$. The
resulting equations are \ba &3(1-\omm)(1+w_0)+g_2=\Phi_1^2 >0&
\label{wc1} \\ &3(1-\omm)((1+w_0)(3\omm w_0 -2)+w_1)-& \nn
\\ &-g_2(4+3(1-\omm)w_0)-g_3=(\Phi^{'2})'(z=0)&
\label{wc2} \\& \frac{3}{2}
(1-\omm)(3(1-w_0^2)-w_1)+\frac{5}{2}g_2+\frac{g_3}{2}> 0&
\label{wc3} \ea In the MCQ limit ($g_2 \rightarrow 0$,
$g_3\rightarrow 0$) equation (\ref{wc3}) has also been obtained in
Ref. \cite{Scherrer:2005je} as a generic limit of MCQ. Using now
(\ref{wc1})-(\ref{wc3}) along with the $w_0-w_1$ $\chi^2$ contours
obtained from the SNLS dataset \cite{Astier:2005qq,Nesseris:2005ur}
we construct Fig. 2 which is a mapping of Fig. 1 on the $w_0-w_1$
parameter space. An interesting point of Fig. 2 is that for $g_2>0$
(Fig. 2a) the forbidden sector I shrinks significantly compared to
MCQ (Fig. 2b) and allows for a $w_0<-1$.

A final set of parameters we consider is the set of the expansion
coefficients of the luminosity distance $d_L(z)$ which in a flat
universe is connected to the Hubble expansion history $H(z)$ as \be
H(z)^{-1}=\frac{d}{dz}\left(\frac{d_L(z)}{1+z}\right) \label{hdl}
\ee Expanding $d_L(z)$ as \be d_L(z)=z  + d_{L2} z^2 + d_{L3} z^3 +
... \label{dlexp} \ee and using the expansion (\ref{hexp}) in
equation (\ref{hdl}) we may express the coefficients $h_1-h_2$ in
terms of $d_{L2}-d_{L3}$ as follows \ba h_1&=&4(1-d_{L2})
\label{hidli1} \\ h_2 &=& 6(1-3d_{L2}+2d_{L2}^2-d_{L3})
\label{hidli2} \ea Substituting now equations (\ref{hidli1}),
(\ref{hidli2}) in equations (\ref{hc1}), (\ref{hc2}) and (\ref{hc3})
we obtain the sector equations in $d_{L2}-d_{L3}$ space as \ba
&4(1-d_{L2})-3\omm +g_2 = \Phi_1^2 > 0 &\label{dlc1} \\
&4d_{L2}(2d_{L2}+g_2-3\omm)+9\omm -& \nn \\ &- 8 -12 d_{L3}-5 g_2 -
g_3 =(\Phi^{'2})'(z=0) &\label{dlc2} \\
&4+8d_{L2}-12d_{L2}^2+6d_{L3}-\frac{9}{2} \omm +& \nn \\
&+\frac{5}{2} g_2 + \frac{g_3}{2} >0& \label{dlc3} \ea Using now
equations (\ref{hidli1}), (\ref{hidli2}) to translate the $h_1-h_2$
$\chi^2$ contours to the $d_{L2}-d_{L3}$ parameter space and
equations (\ref{dlc1})-(\ref{dlc3}) to construct the sectors I-IV in
the $d_{L2}-d_{L3}$ space we obtain Fig. 3. The advantage of Fig. 3
compared to Figs. 1 and 2 is that it refers to the parameters
$d_{L2}-d_{L3}$ which are directly observable through the luminosity
distances of SnIa without the need of any differentiation.

\section{Conclusion-Discussion-Outlook}

We have used generic theoretically motivated inequalities to
investigate the space of observable cosmological expansion
parameters that admits viable MCQ and EXQ theoretical models. Our
inequalities are generic in the sense that they are independent of
the specific features of any scalar field potential (eg scale,
tracking behavior etc) and they only require that the models are
internally consistent. The derived forbidden sectors which violate
the above inequalities already have significant overlap with the
parameter space which is consistent with observations at the
$2\sigma$ level. This overlap which depends on the time derivatives
of the Newton's constant $G$ has lead to a useful constraint to the
second derivative of $G$ (equation (\ref{ddgc})) which is
significantly more stringent compared to the corresponding
constraint (\ref{dglim3}) anticipated from solar system gravity
experiments .

An important reason that limits the observable parameter space
consistent with MCQ and EXQ is the fact that the scalar field
potential energy can induce accelerating expansion but not beyond
the limit corresponding to the cosmological constant ($w(z)=-1$)
obtained when the field's evolution is frozen. Additional
acceleration (superacceleration) can only be provided in the context
of EXQ through the time variation of Newton's constant $G$. A
decreasing $G(t)$ with time favors accelerating expansion and allows
for superacceleration. This physical argument is reflected in our
results. We found that given ${\dot G}_0\simeq 0$ (neglecting $g_1$
since $\vert g_1 \vert < 10^{-4}$ from solar system tests) a ${\ddot
G}_0>0$ ($g_2>0$) decreases the forbidden sector and allows
superacceleration ($w_0<-1$). But ${\dot G}_0\simeq 0$ with ${\ddot
G}_0>0$ implies that we are currently close to a minimum of $G(t)$
with $G(t)$ being larger in the past ie \be
\frac{G(t)}{G_0}\simeq1+\frac{1}{2}g_2 (H_0 (t-t_0))^2 \label{gtev}
\ee Therefore a decreasing $G(t)$ corresponds  to ${\ddot G}_0>0$
($g_2>0$) (see Fig. 4) which in turn implies smaller forbidden
sectors and allows for superacceleration in agreement with the above
physical argument.

\begin{figure}
\rotatebox{0}{\hspace{0pt}\resizebox{.48\textwidth}{!}{\includegraphics{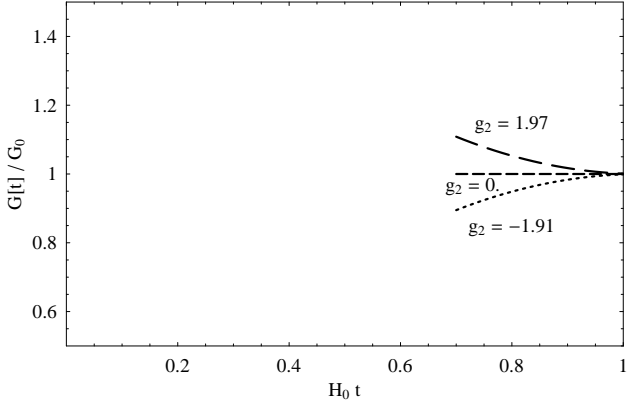}}}
\vspace{0pt}\caption{A $g_2>0$ implies a decreasing Newton's
constant ($\frac{G(t)}{G_0}\simeq1+\frac{1}{2}g_2 (H_0 (t-t_0))^2$
around $t=t_0$).} \label{fig6}
\end{figure}

An additional interesting point is related to the construction of
the $\chi^2$ contours of Figs. 1-3. The SNLS data analysis
involved in the construction of these contours did not take into
account the possible evolution of SnIa due to the evolving $G(t)$.
It is straightforward to take into account the evolution of $G$ in
the SnIa data analysis along the lines of Ref.
\cite{Nesseris:2006jc}. In order to test the sensitivity of these
contours with respect to the evolution of $G$ we have repeated the
$\chi^2$ contour construction assuming a varying $G$ according to
the ansatz \be G(z)=G_0 (1+\alpha \frac{z^2}{(1+z)^2})
\label{gevol} \ee which smoothly interpolates between the present
value of $G=G_0$ and the high redshift value $G=G_0 (1+\alpha)$
implying that \be \alpha=\frac{\Delta G}{G_0} \label{ag} \ee The
parametrization (\ref{gevol}) is consistent with both the solar
system tests\cite{pitjeva} ($\frac{{\dot G}_0}{G_0}\simeq 0$) and
the nucleosynthesis constraints\cite{Copi:2003xd} \be \vert
\frac{G_{nuc}-G_0}{G_0}\vert< 0.2 \label{nucg} \ee at $1\sigma$,
for $\vert \alpha \vert < 0.2$.

It is straightforward to evaluate $g_1,g_2,g_3$ in terms of $\alpha$
using the parametrization (\ref{gevol}) and equation (\ref{tz}). The
result is \ba &g_1 =0& \label{g1a}\\ &g_2  = 2 \alpha
&\label{geffp2}
\\&g_3=3\alpha\left(-1+3 w_0 (-1+\omm )\right)& \label{g3a} \ea We
have considered the value of \be \alpha = 0.2 \label{alv} \ee and
repeated the SNLS data analysis taking into account the evolution
of the SnIa absolute magnitude $M$ due to the evolving $G$ 
\cite{Bond:1984sn,Colgate:1966ax,Amendola:1999vu,Nesseris:2006jc}
as \be M=M_0+\frac{15}{4} log\frac{G}{G_0} \label{abmag} \ee The
steps involved in this analysis may be summarized as follows:
\begin{itemize}\item Use the following magnitude redshift relation
to fit to the SnIa data \be m_{th}(z) = M_0 + 5 log d_L (z) +
\frac{15}{4} log\frac{G(z)}{G_0} \label{magred} \ee where $G(z)$ is
given by (\ref{gevol}) and $d_L(z)$ is connected to $H(z)$ in the
usual geometrically defined way (\ref{hdl}) ie \be d_L (z) =
(1+z)\int_0^z dz' \frac{1}{H(z')} \label{dlh} \ee \item Use an
$H(z)$ parametrization that incorporates the evolution of $G$ (given
by (\ref{gevol})) in the context of the CPL parametrization
(\ref{hcpl}) ie \ba H^2 (z)&=&\frac{G(z)}{G_0} H_0^2 [ \Omega_{0m}
(1+z)^3 + \nn
\\ &+& (1-\Omega_{0m})(1+z)^{3(1+w_0+w_1)}e^{\frac{-3w_1 z}{(1+z)}}]
\label{hcplgev} \ea \item Minimize the $\chi^2$ expression \be
\chi^2 (w_0,w_1)= \sum_{i}^{} \frac{(m^{obs}(z_i) -
m^{th}(z_i;w_0,w_1))^2}{\sigma_{m^{obs}(z_i)}^2} \label{chi2} \ee
and find the corresponding $1\sigma$ and $2\sigma$ $\chi^2$ contours
along the lines of Refs.
\cite{Nesseris:2005ur,Nesseris:2006jc}\end{itemize} The resulting
$\chi^2$ contours in the $w_0-w_1$ space are shown in Fig. 5 (dashed
lines) superposed with the corresponding contours constructed by
neglecting the evolution of $G$ (continuous lines).
\begin{figure}
\rotatebox{0}{\hspace{0pt}\resizebox{.4\textwidth}{!}{\includegraphics{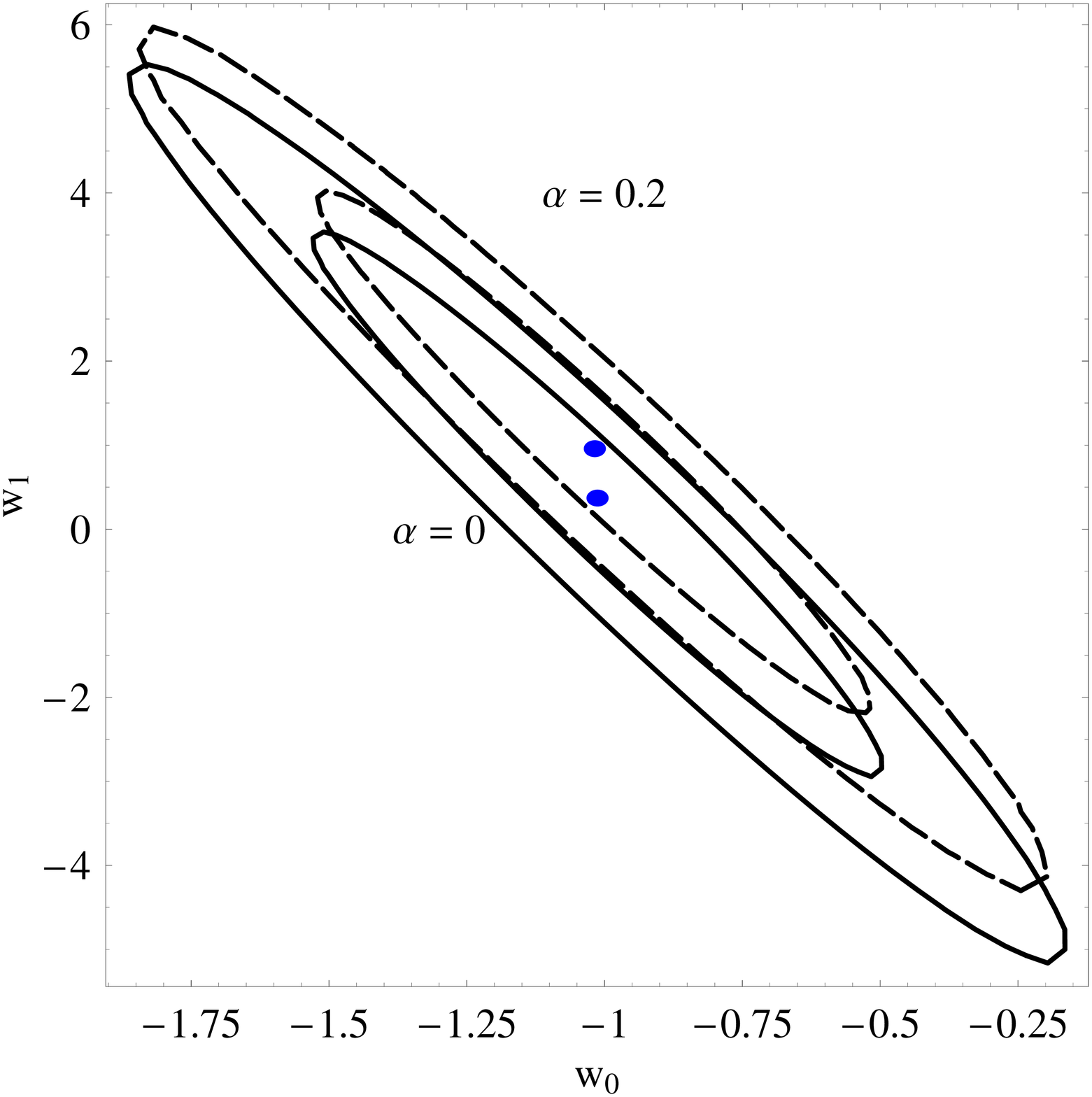}}}
\vspace{0pt}\caption{Taking into account a $G$ variation
consistent with the nucleosynthesis and solar system bounds
introduces modifications to the $w_0-w_1$ $\chi^2$ contours which
are not significant for the currently available SnIa data (for the
dashed line $\alpha=0.2$ and the continuous $\alpha=0$).}
\label{fig5}
\end{figure}
The change of the best fit $w_0-w_1$ values and of the corresponding
errorbars is minor (especially in the $w_0$ direction) and such that
our main conclusion regarding the limiting values of $g_2$ remain
practically unchanged even after the evolution of $G$ is taken into
account in the analysis. This justifies neglecting the evolution of
$G$ in the construction of the $\chi^2$ contours for our purposes.
However, as the quality of SnIa data improves, it becomes clear from
Fig. 5 that the effects of an evolving $G$ {\it consistent with
nucleosynthesis and solar system constraints} on the data analysis
can be significant! This possibility is further amplified if there
are additional effects of an evolving $G$ on the SnIa data analysis.
For example \cite{linder-www,smchdep} it is possible that the time
scale stretch factor $s$ involved in the SnIa data analysis
\cite{Astier:2005qq,Nesseris:2005ur} and arising from opacity
effects in the stellar atmosphere may have a dependence on the
Chandrasekhar mass and therefore on Newton's constant $G$. We have
shown however that even if such effects are included in the SnIa
data analysis our results of Fig. 5 (dashed line) do not change more
than $10\%$.

An interesting extension of the present work could involve the
investigation of the limits of other dark energy models like
braneworld models or barotropic fluid dark energy in the context of
scalar-tensor theories (thus extending the results of Scherrer
\cite{Scherrer:2005je}). The identification of the forbidden
observational parameter regions for such models could be combined
with the constraints of local gravitational experiments to test the
internal consistency of these models. Alternatively the present work
could be extended in the direction of finding the limits and
cosmological consistency of specific classes of potentials in the
context of EXQ thus also extending the work of
\cite{Sahlen:2006dn,Huterer:2006mv,Li:2006ea}.

{\bf Numerical Analysis:} Our numerical analysis was performed using
a Mathematica code available at {\bf
http://leandros.physics.uoi.gr/exqlim/exqlim.htm }
\section*{Acknowledgements}
We thank J. Muller for providing the preliminary unpublished solar
system result for $\frac{{\ddot G}_0}{G_0}$ shown in equation
(\ref{mulbound}). We also thank D. Polarski, A. Starobinsky and E.
Linder for useful comments. This work was supported by the
European Research and Training Network MRTPN-CT-2006 035863-1
(UniverseNet). SN acknowledges support from the Greek State
Scholarships Foundation (I.K.Y.).

\end{document}